\documentclass[a4paper]{article}

\usepackage{INTERSPEECH2020}
\usepackage{graphicx}
\usepackage{url}

\title{X-Vectors with Multi-Scale Aggregation for Speaker Diarization}
\name{Myungjong Kim, Vijendra Raj Apsingekar, Divya Neelagiri}
\address{
  Samsung Research America, Mountain View, USA}

\email{myungjong.k@samsung.com, v.akar@samsung.com, d.neelagiri@samsung.com}

\begin{document}

\maketitle
\begin{abstract}
Speaker diarization is the process of labeling different speakers in a speech signal. Deep speaker embeddings are generally extracted from short speech segments and clustered to determine the segments belong to same speaker identity. The x-vector, which embeds segment-level speaker characteristics by statistically pooling frame-level representations, is one of the most widely used deep speaker embeddings in speaker diarization. Multi-scale aggregation, which employs multi-scale representations from different layers, has recently successfully been used in short duration speaker verification. In this paper, we investigate a multi-scale aggregation approach in an x-vector embedding framework for speaker diarization by exploiting multiple statistics pooling layers from different frame-level layers. Thus, it is expected that x-vectors with multi-scale aggregation have the potential to capture meaningful speaker characteristics from short segments, effectively taking advantage of different information at multiple layers. Experimental evaluation on the CALLHOME dataset showed that our approach provides substantial improvement over the baseline x-vectors.

\end{abstract}

\section{Introduction}

Speaker diarization is the process of labeling different speakers in an audio stream, responding the question ``who spoke when'' in a multi-speaker conversation \cite{anguera2012speaker}. It has been received great attention due to the potential in a variety of applications such as meeting conversation analysis and multi-media information retrieval. For meeting conversation analysis, in particular, speaker diarization can be used as a front-end component of automatic speech recognition (ASR), providing improved ASR accuracy and more rich analysis depending on participants.

General speaker diarization systems have several steps \cite{anguera2012speaker, sell2014speaker}. First, speech activity detection is applied to filter out non-speech frames and the resulting speech segments are divided into short segments. Second, speaker embeddings are extracted from the short speech segments. Third, the extracted embeddings are clustered to determine the segments belong to same speaker identity. 

One of the biggest challenges of speaker diarization is to extract speaker embeddings from short speech segments. Traditionally, i-vectors were widely used in speaker diarization as a feature vector from segments \cite{sell2014speaker, shum2013unsupervised}. Recently, deep speaker embedding approaches are introduced and widely used in speaker diarization and short-duration speaker verification. In deep speaker embedding approaches, deep neural networks are trained to represent speaker identity using cross-entropy loss \cite{snyder2018x, garcia2017speaker}, or triplet loss \cite{song2018triplet, li2017deep} and the representations at the last hidden layer are used as speaker embeddings. Feed-forward deep neural networks (DNNs) \cite{garcia2017speaker}, convolutional neural networks (CNNs) \cite{cyrta2017speaker}, and long short-term memory recurrent neural networks (LSTM-RNNs) \cite{wan2018generalized, wang2018speaker} were used as deep speaker embedding networks. To allow variable length segment embeddings, frame-level representations from the networks are aggregated over time using a pooling layer such as global average pooling \cite{li2017deep} or statistics pooling \cite{snyder2018x}. Particularly, x-vectors \cite{snyder2018x} embed segment-level speaker characteristics by statistically pooling frame-level representations from time-delay neural networks (TDNNs) and have been successfully proven its effectiveness in speaker diarization challenges \cite{sell2018diarization}. 

Conventional speaker embedding networks use a pooling layer from representations of the last layer of frame-level neural networks. Since the pooling layer receives single-scale representations from one layer, it might not capture useful speaker information at other layers. Multi-scale aggregation (MSA) has recently been introduced to aggregate multiple scale representations from different layers of frame-level neural networks and shown improved accuracy in speaker verification tasks \cite{jung2020improving}. In \cite{gao2019improving}, multiple feature maps from different layers were combined as a single feature map and pooling operation was applied to the feature map to represent utterance-level speaker characteristics. In \cite{seo2019shortcut},  pooling operation was applied to multiple feature maps from different layers and the features were concatenated for speaker representation. Hajavi and Etemad also used multiple embeddings from different layers to generate short-segment-level speaker representations \cite{hajavi2019deep}. Particularly, Tang \emph{et al.} used two statistics pooling layers from different two frame-level layers in x-vector embedding networks for speaker verification tasks \cite{tang2019deep}. Speaker diarization with short speech segments requires maximizing the use of the limited amount of information within the segment \cite{hajavi2019deep}. Therefore, the MSA approaches might be more effective in speaker diarization tasks. However, it has rarely been studied in speaker diarization.

In this paper, we investigate an MSA approach in an x-vector embedding framework for speaker diarization. To this end, we first explore several frame-level neural network architectures such as standard TDNNs \cite{peddinti2015time}, extended TDNNs (E-TDNNs) \cite{snyder2019speaker}, and factorized TDNNs (F-TDNNs) \cite{povey2018semi}. After that, an MSA scheme is proposed based on F-TDNNs. Our approach was evaluated on the CALLHOME dataset and the proposed method provided better speaker diarization accuracy than the baseline x-vectors. 

The rest of the paper is organized as follows: Section 2 describes our speaker diarization system with single-scale x-vector embedding networks. Section 3 discusses our proposed multi-scale aggregation in an x-vector framework. Experimental setup and results are presented in Section 4 and 5, respectively. Finally, the conclusions are given in Section 6. 


\section{Speaker Diarization with X-Vectors}

This section describes the x-vector based speaker diarization method. Given an audio recording, we extract 23 dimensional mel-frequency cepstral coefficients (MFCCs) with a frame size of 25 ms and a shift size of 10 ms as frame-level features and the features are mean-normalized over a sliding window of up to 3 seconds. After non-speech frames are filtered out, the speech segments are divided into sub-segments, e.g., 1.5 seconds long with 0.75 seconds overlap. The x-vector speaker embeddings are extracted from the sub-segments. The extracted x-vectors are whitened using principal component analysis (PCA) transforms and further processed by applying conversation-dependent PCA transforms to reduce the dimensions (will be discussed further in Section 4.2). Probabilistic linear discriminant analysis (PLDA) scores are calculated between all pairs of transformed x-vectors. Finally, the PLDA scores are clustered using agglomerative hierarchical clustering  (AHC) \cite{sell2014speaker}. In this work, we explore the following three x-vector embedding networks: TDNN, E-TDNN, and F-TDNN.

\subsection{TDNN-based x-vectors}

The x-vector \cite{snyder2018x} embedding networks are composed as follows: 1) frame-level layers that handle speech frames, 2) a statistics pooling layer that aggregates the frame-level representations over a segment, 3) additional fully connected layers that operate at the segment-level (embeddings are obtained here), and 4) finally a softmax output layer. 

The TDNN-based x-vector network architecture is same as in Kaldi recipe \cite{snyder2018callhome, povey2011kaldi} except the embedding layer size\footnote{The best performance was obtained at 512 dimension in our experiments. Thus, we used 512 in this work.}. The first 5 frame-level layers of the network are based on TDNNs, which is a special type of 1-D CNNs. Suppose $t$ is the current time step. At the input, the frames are spliced at  $\{t-2, t-1, t, t+1, t+2\}$. For the next two layers, the output of the previous layer is spliced at times $\{t-2, t, t+2\}$ and $\{t-3, t, t+3\}$, respectively. The next two layers are also frame-level layers, but without any added temporal context. The number of hidden units at each hidden layer are 512 except fifth layer with a size of 1500. 

The statistics pooling layer aggregates the output of the final frame-level layer over the input segment (i.e., 150 frames) and computes its mean and standard deviation, resulting in a size of 3000. The segment-level statistics are connected with one hidden layer with a size of 512, which will be an x-vector, and finally the softmax output layer with a size of the number of speakers.

\begin{table}[t]
  \caption{F-TDNN architecture}
  \label{tab:word_styles}
  \centering
  \scalebox{0.72}{
  \begin{tabular}{cccccc}
    \toprule
    \textbf{Layer}      & \textbf{Layer Type}   & \textbf{Context}  & \textbf{Skip conn.} & \textbf{Size}   & \textbf{Inner}      \\
         &   &  & \textbf{from layer} &    &   \textbf{size}    \\
    \midrule
    frame1   & TDNN-ReLU   &  t-2:t+2         &    &    512 &                    \\
    frame2   & F-TDNN-ReLU    & t-2, t, t+2 &    &   725 &      180                \\
    frame3   & F-TDNN-ReLU    & t                &     &   725 &     180 \\
    frame4   & F-TDNN-ReLU    & t-3, t, t+3 &     &   725 &     180            \\
    frame5   & F-TDNN-ReLU    & t                 & frame3  &  725 &     180   \\
    frame6   & F-TDNN-ReLU    & t-3, t, t+3  &    &   725  &    180   \\
    frame7   & F-TDNN-ReLU    & t-3, t, t+3  &   frame4  &  725 &   180  \\
    frame8   & F-TDNN-ReLU    & t-3, t, t+3  &    &     725 &   180               \\
    frame9   & F-TDNN-ReLU    & t                  &  frame6 & 725 &     180 \\
    frame10  & Dense-ReLU       & t                  &   &  1500 &   \\
    seg.11   & Pooling (mean+stddev) & segment &  & 2 x 1500 & \\
    seg.12   & Dense-ReLU       &     &   & 512 &   \\
    seg.13   & Dense-ReLU        &   &  & 512 &  \\
    seg.14   & Dense-Softmax      &   &   & No. spks &                  \\
    \bottomrule
  \end{tabular}}
  \vspace{-1mm}
\end{table}

\subsection{Extended TDNN-based x-vectors}

The extended TDNN (E-TDNN) uses a similar structure in Section 2.1, but wider temporal context in the TDNN layers, and interleave dense layers between the TDNN layers. We followed the E-TDNN architecture in  \cite{snyder2019speaker}. The first 10 layers are frame-level layers with context $\{t-2, t-1, t, t+1, t+2\}$, $\{t-2, t, t+2\}$, $\{t-3, t, t+3\}$, and $\{t-4, t, t+4\}$ at first, third, fifth, and seventh layer, respectively. Other layers are dense layers without any added temporal context. The pooling layer receives the output of layer 10 as input. There are two more segment-level dense layers followed by the softmax output layer. The last hidden layer is also considered as an x-vector. The numbers of hidden units for the all frame-level and segment-level layers are 512 except the layer 10 with a size of 1500.  

\begin{figure}[t]
  \centering
  \scalebox{0.97}{
  \includegraphics[width=\linewidth]{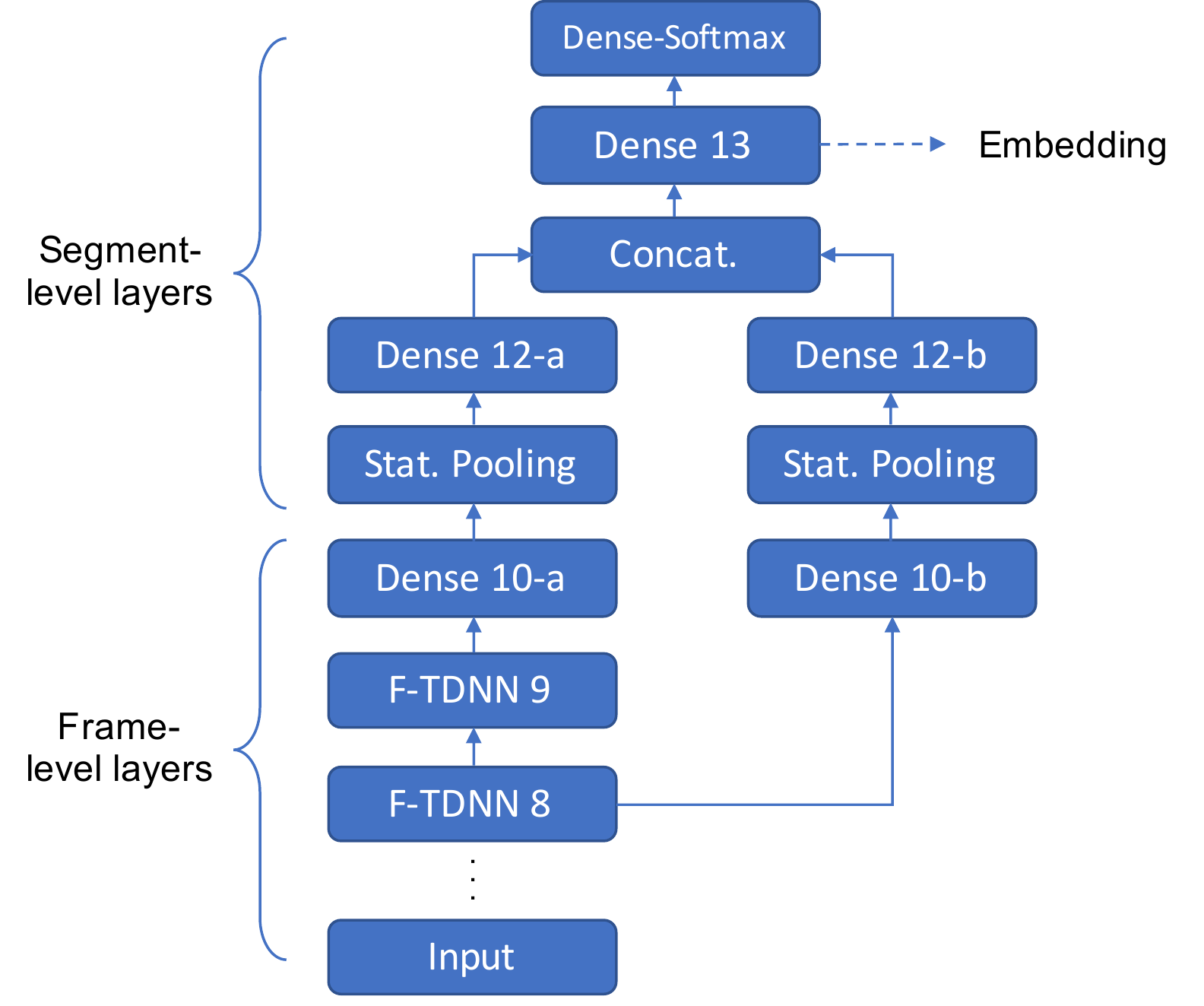}}
  \caption{Proposed x-vector embedding network with multi-scale aggregation.}
  \label{fig:xvec_msa}
  \vspace{-1mm}
\end{figure}

\subsection{Factorized TDNN-based x-vectors}

The factorized TDNN (F-TDNN) \cite{povey2018semi} is a factored form of TDNN. In other words, the weight matrix of each TDNN layer is factorized into the product of two low-rank matrices. The first of those factors is constrained to be semi-orthogonal. It is expected that the semi-orthogonal constraint helps in not loosing modeling power while the high-dimension vectors are projected into a lower-dimensional feature space, thus aiding in reducing the model size compared to TDNN models.

When factorizing the TDNN layer, the layer is factorized using two convolutions with half the kernel size as in \cite{povey2018semi}. For the context $\{t-2, t, t+2\}$, a kernel with context $\{t-2, t\}$ in the first factor and a kernel with context $\{t, t+2\}$ in the second factor were used. 

To effectively model the deep F-TDNNs by alleviating the vanishing gradient problem, we use skip connections \cite{he2016deep} between adjacent layers as well as other layers. In other words, some frame-level layers receive as input, not only the previous layer but also the output from other prior layers. In \cite{snyder2019jhu}, the prior layers were concatenated to the input of the current layer. In this work, we used element-wise summation instead of concatenation. Table 1 summarizes our F-TDNN architecture. Compared to \cite{snyder2019jhu}, we used less skip connections with element-wise summation and also pooling over short segments. The E-TDNN and F-TDNN were applied in multi-speaker recognition tasks \cite{snyder2019speaker}, which used first speaker diarization followed by speaker recognition. However, the diarization accuracy was not measured. In \cite{huang2018jhu}, the F-TDNN was applied for speaker diarization, but specific architectures were not described.

\section{X-Vectors with Multi-Scale Aggregation}

Conventional x-vectors use one pooling layer to gather frame-level representations from the last frame-level hidden layer. Although the last hidden layer has the most important high-level speaker characteristics at frame-level, it might not capture useful speaker information at other prior layers. Multi-scale aggregation (MSA) can alleviate these problems by utilizing multiple scale representations from different frame-level layers. Speaker representation with short speech segments needs to maximize the use of the limited amount of information within the segment \cite{hajavi2019deep}. Thus, MSA in an x-vector embedding network might provide better modeling capability to represent speaker characteristics in a short segment for speaker diarization. 

Our proposed x-vector embedding networks with MSA is shown in Figure 1. This network is based on the F-TDNN architecture described in Section 2.3. Compared to the single-scale F-TDNN, we have another statistics pooling layer from the 8th F-TDNN layer (i.e., F-TDNN 8 in Figure 1). Specifically, the 8th F-TDNN layer is connected to the dense layer (i.e., Dense 10-b) and statistics pooling operation is exploited through the output of the dense layer. The pooling output is linked with another dense layer (i.e., Dense 12-b). We also keep the original path from 9th F-TDNN layer (i.e., F-TDNN 9) to statistics pooling and segment-level dense layers (i.e., Dense 12-a). Finally, the outputs of dense layers from two statistics pooling layers are concatenated and fed into another dense layer (Dense 13). The output of Dense 13 is the embedding vector.

The number of dimensions at each frame-level F-TDNN layer is same as in Table 1, which means F-TDNN 8 has 725 hidden units with an inner linear bottleneck size of 180. The frame-level dense layers (Dense 10-a and 10-b) have 1500 hidden units, respectively. The first segment-level dense layer (Dense 12-a and 12-b) have 256 hidden units, respectively. The last segment-level dense layer (the embedding layer) has 512 dimensions.

\section{Experimental Setup}

\subsection{Training data}

The training data includes SWBD, SRE, and Voxceleb \cite{Nagrani17, Chung18b}  datasets. The SWBD set consists of Switchboard 2 Phase 2 $\&$ 3 and Switchboard Cellular 1 $\&$ 2. In total, the SWBD set contains about 21k recordings from 1991 speakers. The SRE set is 2004 NIST SRE set and contains 4.5k recordings from 307 speakers. The Voxceleb dataset consists of Voxceleb 1 training set $\&$ 2 and contains 1.2m recordings from 7323 speakers. All the Voxceleb data is downsampled to 8kHz. In addition, data augmentation is applied to double the size and diversity of the training set. The augmentation is done using additive noises (5-15dB SNRs) from the MUSAN dataset \cite{musan2015} and reverberation based on the room impulse responses (RIRs) dataset\footnote{https://openslr.org/28/}. To augment a recording, we choose between one of the following randomly: background babble, background music, foreground noise, reverberation. Finally, 1m augmentation data was randomly chosen and was used in addition to the original set. 

\subsection{X-vector and PLDA model training}

The x-vector embedding networks are trained on SWBD + SRE + Voxceleb + augmentation and the PLDA transforms are trained on SRE. To train the neural networks, we excluded short utterances that are less than 5 seconds long and any speaker with fewer than 8 utterances in the training set. It results in 9230 speakers with 1.5m recordings including augmentation data. 

To efficiently train the network, we randomly split the recordings into examples that range from 2 to 4 seconds along with a minibatch size of 64. It results in 3,200 examples per speaker. The loss function is a multi-class cross entropy loss function to predict speakers from the examples. In our x-vector network, the segment size is 1.5s, and therefore, we used sliding window averaging within the examples. The network is trained from 3 to 5 epochs using the natural gradient stochastic gradient descent algorithm \cite{povey2014parallel}. At each hidden layer, the ReLU activation and batch normalization are used. Also, we regularize the network using dropout and L2 regularization.

After training the embedding network is done,  x-vectors are extracted from the training set, length-normalized, and used to train a PCA whitening matrix. Based on our preliminary experiments, we decided to keep 512 dimensions after the PCA transform for single-scale x-vectors while x-vectors with MSA reduced to 150 dimensions using the transform.  The transformed x-vectors are used to train a PLDA matrix. 

Within each conversation in testing, we compute a conversation-dependent PCA \cite{sell2014speaker} to further reduce the dimension after x-vectors are length-normalized and transformed using the whitening matrix to preserve 10\% of dimensions . Similarly, the parameters of the PLDA matrix were projected using the conversation-dependent PCA matrix. The PLDA scores are calculated between all x-vector pairs and clustered using AHC.

\subsection{Evaluation data and metric}

Our algorithms are evaluated on the CALLHOME dataset that consists of 500 conversations of between 2 and 7 speakers\footnote{https://openslr.org/10/sre2000-key.tar.gz}. Within each conversation, all speakers are recorded in a single channel. 

The evaluation measure is diarization error rates (DERs), which combines all types of error such as missed speech, mislabeled non-speech, and incorrect speaker cluster. However, we used oracle speech activity detection (SAD) marks, and therefore, only incorrect speaker labeling factors are considered as the DER. Also, DER tolerated errors with 250ms of a speaker transition and ignored overlapping segments in scoring. 

We report DERs on two conditions with/without the actual number of speakers in conversations. The former uses the actual speaker number during clustering (referred as \emph{Oracle} in Table 2 and 3) and for the later, we find the optimal threshold for clustering in a supervised manner using 2-fold cross validation. In other words, we split the CALLHOME set into two sets and uses one set for the development to find the best performed threshold and the other set for testing given threshold (referred as \emph{Threshold}).

\section{Results}

\subsection{Effect of F-TDNN x-vectors}

\begin{table}[t]
  \caption{DERs (\%) of different x-vector models}
  \label{tab:word_styles}
  \centering
  \begin{tabular}{ccc}
    \toprule
    \textbf{Model}      & \textbf{Threshold}      & \textbf{Oracle}           \\
    \midrule
    TDNN (Section 2.1)                  & 9.07                     & 7.96                  \\
    E-TDNN (Section 2.2)                & 9.05                  & 7.89                  \\
    F-TDNN (concatenation)            & 8.54 & 7.89                          \\
    F-TDNN (addition from prev. layer)                 & 8.58 & 7.57                               \\
    F-TDNN (Section 2.3)        & \textbf{8.45} & \textbf{7.42}                   \\

    \bottomrule
  \end{tabular}
  \vspace{-1mm}
\end{table}

Table 2 shows the DERs of different single-scale x-vectors. As shown in Table 2, the F-TDNN based x-vectors present better performance than the TDNN- and E-TDNN-based x-vectors for both the threshold and oracle settings. For the F-TDNN models, we compared concatenation-based skip connection from multiple layers, element-wise addition-based skip connection from only previous layer, and from multiple layers. Our F-TDNN-based x-vector model based on element-wise addition from multiple layers was the best on the single-scale x-vector comparison, providing a DER of 8.45\% for the threshold and a DER of 7.42\% for the oracle setting.

\subsection{Effect of multi-scale aggregation}

\begin{table}[t]
  \caption{DERs (\%) of x-vectors with MSA}
  \label{tab:word_styles}
  \centering
  \begin{tabular}{ccc}
    \toprule
    \textbf{Model}      & \textbf{Threshold}      & \textbf{Oracle}           \\
    \midrule
    F-TDNN                   & 8.45                     & 7.42                  \\
    F-TDNN  (stat: 6000 dim)                 &        8.79            &       8.05            \\
    \midrule
    F-TDNN-MSA (8 $\&$ 9)                & \textbf{8.00}                  & \textbf{7.25}                  \\
    F-TDNN-MSA (7 $\&$ 9)            & 8.35 & 7.71                          \\
    F-TDNN-MSA (7 $\&$ 8 $\&$ 9)                 &  8.42  &     7.43                           \\

    \bottomrule
  \end{tabular}
  \vspace{-1mm}
\end{table}

Table 3 shows the diarization results of x-vectors with MSA. Since the MSA method uses two pooling layers with same size in F-TDNN, we also double the size of statistics pooling layer (i.e., from 3000 to 6000) in single-scale F-TDNN for the fair comparison in terms of model complexity. For the F-TDNN-MSA approaches, we also tested the MSA connections from the 7th $\&$ 9th layers and 7th $\&$ 8th $\&$ 9th layers. As shown in Table 3, the best DER can be found on F-TDNN-MSA described in Section 2.4, giving a DER of 8.00\% and 7.25\% for both conditions, respectively. 

We also compared the diarization performance depending on the number of speakers in conversations in Figure 2. Here, we report DERs obtained on the threshold setting. In the CALLHOME dataset, two or three speakers are dominant, and therefore, we grouped four or more speakers. As can be seen, the proposed F-TDNN-MSA gives the best results for all cases (4.98\%, 6.37\%, and 13.08\% DERs for 2, 3, and above 4 speakers.  

\begin{figure}[t]
  \centering
  \includegraphics[scale=0.53]{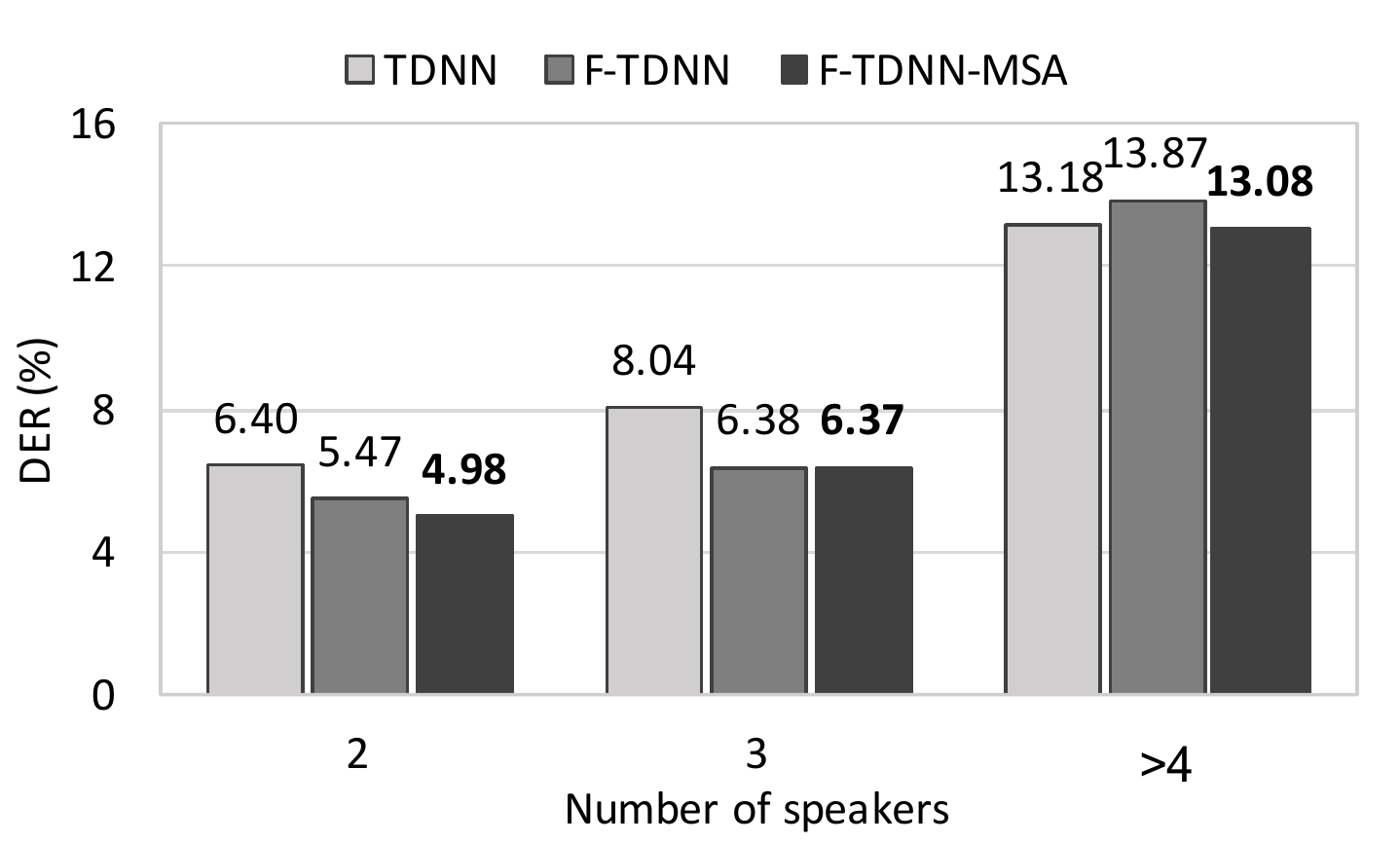}
  \caption{DERs depending on the number of speakers in conversations.}
  \label{fig:xvec_msa}
  \vspace{-1mm}
\end{figure}


\subsection{Comparison with previous methods}

Table 4 compares our F-TDNN-MSA method with other recent speaker diarization methods reported on the same CALLHOME dataset based on oracle SAD or incorrect speaker labeling as the DER. The inside parenthesis indicates speaker embedding + clustering methods in their works. Here, VBEM-GMM means the variational Bayes expectation-maximization Gaussian mixture model and UIS-RNN is the unbounded interleaved-state RNN. +VB means VB refinement as a post processing. d-vector1, 2, and 3 (which is LSTM-RNN-based embeddings) represent its version depending on training set size and combination.

For the comparison of speaker embeddings, the x-vectors show better results when using widely used clustering methods such as AHC or spectral clustering. Our F-TDNN-MSA x-vectors achieved the state-of-the-art performance on that. Based on the d-vectors and x-vectors, the performance was much improved when combining advanced clustering methods such as UIS-RNN and auto-tuned spectral clustering methods. 

\subsection{Discussion}

We focused on modeling better speaker embedding networks for speaker diarization in this work. We observed that our F-TDNN-MSA x-vectors showed the lowest DERs based on the standard clustering methods, but still performance gap with the state-of-the-art clustering methods. Therefore, our method might have the potential to achieve better accuracy by using the aforementioned advanced clustering methods.

In addition, we also observed the performance gap between our TDNN x-vector and Snyder \& Maciejewski  TDNN x-vector \cite{snyder2018callhome}. One possible reason is the difference on training data. In our training set, the amount of mismatched (wideband) condition data, e.g., Voxceleb, is much larger than with matched (telephone) condition data. Thus, further investigation is needed using same condition data in the future.

\begin{table}[t]
  \caption{Comparison with previous methods}
  \label{tab:word_styles}
  \centering
  \scalebox{0.91}{
  \begin{tabular}{cc}
    \toprule
    \textbf{Method}          & \textbf{DER (\%)}           \\
    \midrule
    Shum \textit{et al.} \cite{shum2013unsupervised}  (i-vector+VBEM-GMM)                                   & 14.5                  \\
    Senoussaoui \textit{et al.}  \cite{senoussaoui2013study}    (i-vector+mean-shift)                     &       12.1            \\
    Sell \textit{et al.} (+VB)    \cite{sell2015diarization} (i-vector+AHC)                            & 13.7 (11.5)                  \\
    Garcia-Romero \textit{et al.} (+VB)   \cite{garcia2017speaker}  (DNN+AHC)       & 12.8 (9.9)                          \\
    Wang \textit{et al.} \cite{wang2018speaker} (d-vector1+spectral)            & 12.0                          \\
     Zhang \textit{et al.} \cite{zhang2019fully}  (d-vector1+UIS-RNN)      & 10.6    \\
    Zhang \textit{et al.} \cite{zhang2019fully}  (d-vector2+UIS-RNN      & 9.6    \\
    Zhang \textit{et al.} \cite{zhang2019fully}   (d-vector3+UIS-RNN)     & 7.6  \\
    \midrule
    Snyder \& Maciejewski   \cite{snyder2018callhome}    (TDNN x-vector+AHC)             & 8.39   \\
    Park \textit{et al.}  \cite{park2019auto} (TDNN x-vector+auto-tune spectral) & 7.25   \\
    \textbf{This work (F-TDNN-MSA x-vector+AHC)}                   &       \textbf{8.00}                        \\

    \bottomrule
  \end{tabular}}
  \vspace{-1mm}
\end{table}

\section{Conclusions}

In this paper, we proposed the x-vector embedding network with MSA for speaker diarization. We used multiple statistics pooling layers from the last two frame-level F-TDNN layers and the pooled representations are concatenated to generate speaker embeddings. A series of experiments was performed in terms of the DER on the CALLHOME dataset. Experimental results showed that the proposed method provides meaningful improvement over the baseline x-vectors. Our approach presents a possibility in effectively modeling speaker embeddings in short speech segments.



\bibliographystyle{IEEEtran}

\bibliography{ms}


\end{document}